# Network state Estimation using Raw Video Analysis: vQoS-GAN based non-intrusive Deep Learning Approach

**Renith G [a], Harikrishna Warrier [b], Yogesh Gupta [c]**

[a] HCL Technologies, India, renith.r@hcl.com

[b] HCL Technologies, India, harikrishna.w@hcl.com

[c] HCL Technologies, India, yogeshg@hcl.com

**Abstract**

Content based providers transmits real time complex signal such as video data from one region to another. During this transmission process, the signals usually end up distorted or degraded where the actual information present in the video is lost. This normally happens in the streaming video services applications. Hence there is a need to know the level of degradation that happened in the receiver side. This video degradation can be estimated by network state parameters like data rate and packet loss values. Our proposed solution vQoS-GAN (video Quality of Service – Generative Adversarial Network) can estimate the network state parameters from the degraded received video data using a deep learning approach of semi supervised generative adversarial network algorithm. A robust and unique design of deep learning network model has been trained with the video data along with data rate and packet loss class labels and achieves over 95 percent of training accuracy. The proposed semi supervised generative adversarial network can additionally reconstruct the degraded video data to its original form for a better end user experience.

1. **Introduction**

Digital streaming services have become more popular nowadays in various domains since it serves to progress the business continuity in a smoother way. There are many popular video streaming services that stream the video data to the end user located anywhere in the world. The video streaming services buffer the video data that has been transmitted over the network encapsulated in smaller multiple packets. The receiver side that receives the transmitted video data will aggregate all the packets that have been received and starts displaying to the end user through the media player. There are also other real time applications where the video streaming services are live. These real time streaming applications were very much useful for many social and commercial events during the recent pandemic situation.

## 1.1. Digital Streaming Service: Social Contribution

Over-the-Top (OTT) video media services and other interactive live video streaming service demands went high during the pandemic situation like Covid-19. These streaming services helped various kinds of people to resume their work without any hindrance. School / College students connected with their respective professors to continue their education in an online mode. Industrial business meetings were conducted regularly, and the progress was monitored in real time to maintain good smoother business continuity. And even people isolated in various countries due to sudden stoppage in transport were able to connect with their family members with these live streaming services. There are many such examples where these streaming services helped during this pandemic period.

## 1.2. Need for Network State Estimation

One common factor to use these OTT and other streaming service is Internet. In today's market, there are multiple OTT and live video streaming services that can be found. When all these service providers use the same internet infrastructure, then there is high possibility of congestion over the network. Due to the network congestion, either there is a chance of data loss during streaming, or the service cannot be properly delivered to the consumer. To avoid these network congestions, there is a requirement to throttle and reduce the video traffic to the end user or consumer. For controlling the video traffic to consumers, the streaming services need to estimate the network state parameters like data rate and packet loss.

## 1.3. Video Quality Degradation

There is no guarantee that the received video data have the same video quality as the original transmitted data. The speed and quality of the streaming services are based on few important network parameters such as data rate and packet loss rate. If these parameters are not properly maintained, then there is a possible chance of higher degradation of video data in the receiver side. If the degradation level affects the visual quality of the video data, then the purpose of streaming the video data goes in vain. Hence there is a need for the service provider to know about the degree of video data degradation in the receiver side. To understand about the degradation level, we need to compare both the original non degraded transmitted video and the received degraded video data. There are some existing traditional techniques using probe method which can compare the data. A brief view of a probe-based method is given in the Figure 1.

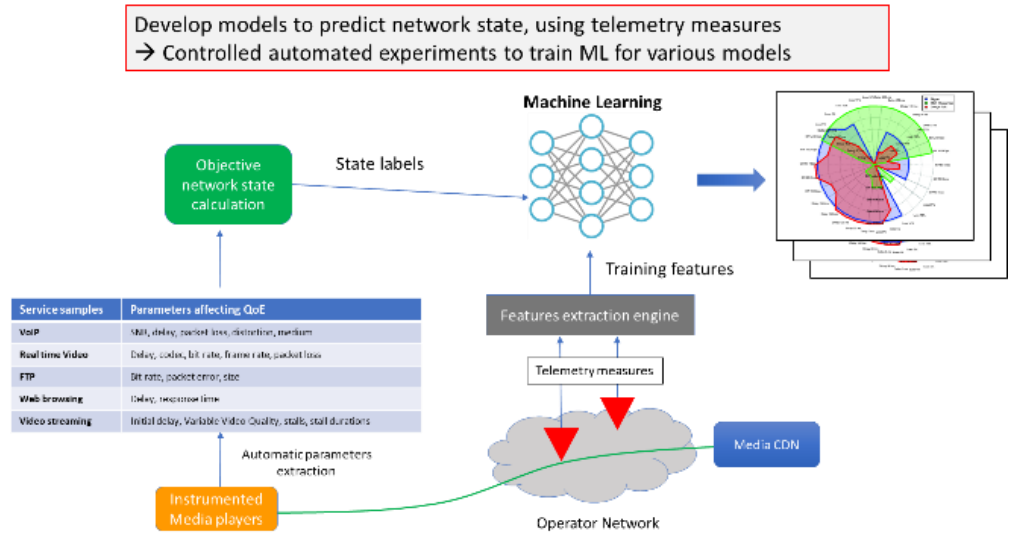

**Figure 1. A sample implementation of a probe-based approach to estimate network state information**

In this method, specialized hardware probes are employed to monitor and record the network traffic in real time and use the data to train a machine learning model for network state estimation.

But in a real-world application, we have the information of degraded video only and it may not be possible to tap into the network at different points to monitor the network traffic degradation. The actual quality of original input video is also unknown and so the service provider cannot determine the degree of degradation. Hence, we need a non-intrusive mechanism, a kind of automation, to estimate the network state parameters from the degraded video frame alone.

**1.4. Data Rate and Packet Loss**

The amount of information or the data that is been transmitted over the network during specific interval of time is termed as Data Rate. The quality of the video in the receiver side mainly depends on the data rate transmission over the network. Higher the data rate implies good quality of video in the receiver side.

During transmission of data over the network, the entire data gets split into various small batches called as Packets. These packets might be lost if the congestion in the network is high. The rate of losing the packet information is termed as packet loss rate. For a minimal data rate video transmission, the impact of packet loss will be heavier.

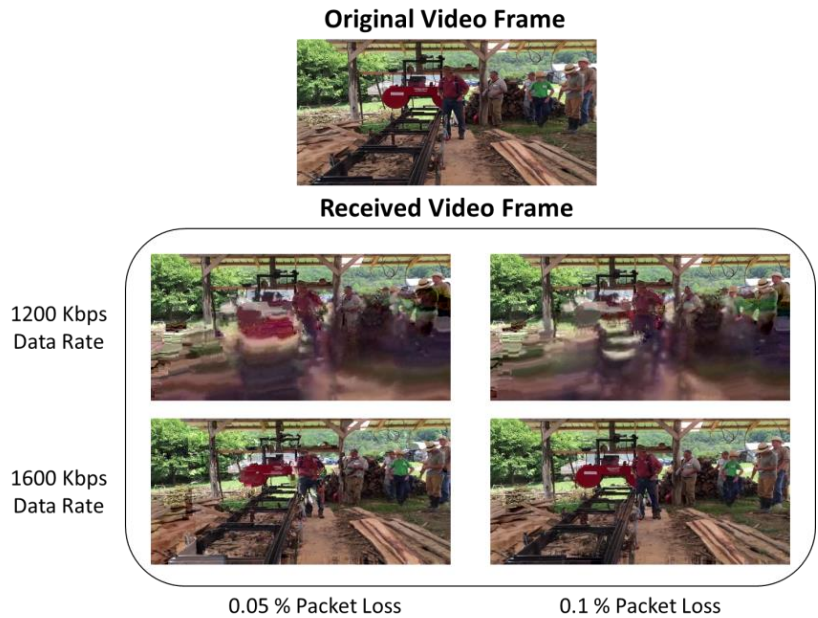

Figure 1. Video Quality Degradation based on Data Rate and Packet Loss

The Quality of video degradation can be clearly seen with respect to Data Rate and Packet loss in Figure 2. For a data rate of 1200 kbps, the degradation is higher based on packet loss of 0.05% and 0.1%. For a higher data rate of 1600 kbps, the degradation is reduced based on the respective packet loss.

For both service provider and end user, there are two solution aspects that need to be addressed in this quality degradation problem. The first aspect should aim in finding the level of degradation that happened in the received video without having knowledge about input original video quality. The second solution aspect should aim in restoring the degraded video data to its original video form to provide a gratifying end user experience. Our proposed solution takes care of both aspects and helps the service provider to find the degradation level in receiver side and for enriching the end user experience by restoring the degraded video to its original video form. Estimating the network state can be automated using robust deep learning algorithms. The network state parameters such as data rate and packet loss are considered for this problem statement. The design is also extendable to estimate other network state parameters like latency, jitter etc. by adding more class labels to the proposed deep learning network architecture and train the model accordingly.

## 2. Related Work

Image and video quality enhancement research has been extensively given more importance since it serves as a core data for many applications. Analysis has been done on recent research papers where the researchers have tried to deblur or enhance the video quality from the degraded video or image data.

Nahli et al [1], has proposed an algorithm called DeblurNet for performing deblurring operation on the degraded image data. The author has achieved by using two modules called frame selection and frame deblurring. Frame selection module finds the blurred input frames and then gives to deblurring module for deblur the degraded data. Zhang et al [2], has proposed an encoder decoder network algorithm for performing the deblur operation in the video data. The video frame needs to be deblurred is selected and then given to the encoding decoding network algorithm.

Liang et al [3], has proposed a deep learning model for deblurring the image data. A unique approach taken is to perform deblurring operation on raw images rather than performing in a RGB image data. Kim et al [4], has proposed a deep learning model in performing deblurring operation for a motion data using convolutional neural network algorithm. Jiang et al [5], has proposed an algorithm for deblurring motion data using convolutional recurrent neural network by integrating both visual knowledge and temporal knowledge of the input data. Su et al [6], has proposed a convolutional neural network deep learning algorithm for improving the video data quality as an end-to-end solution.

Tong et al [7], has proposed a deep learning solution for enhancing the video quality using custom convolutional neural network algorithm. Applying an optical flow method for studying the successive neighboring frames and then feeding to a deep neural network gives good result in enhancing the video quality. Huang et al [8], has proposed a deep learning technique for enhancing the quality for intra frames present in the input video data. Khasanova et al [9] has proposed a denoising algorithm using deep learning technique for enhancing the video quality in the surveillance applications.

## 3. Dataset Description

The public video dataset taken from ITU AI/ML 5G Network state estimation has been used for creating a proof of concept of our proposed solution. The dataset comprises of both original non degraded video files and its corresponding received degraded video files. Each original video file has a combination of 10 different data rates and 6 different packet loss rates of received degraded video files. The received degraded video frames are created in a lab environment using the network emulator with predefined data rate and predefined packet loss. For the analysis purpose, a sample original non degraded video file and its corresponding four different received degraded video files are taken with combination of two different data rate and two different packet loss value. The sample original non degraded video frame and received degraded video frame from the ITU AI/ML dataset is shown in Figure 2. This shows how the degraded video file differs from the actual original non degraded video file.

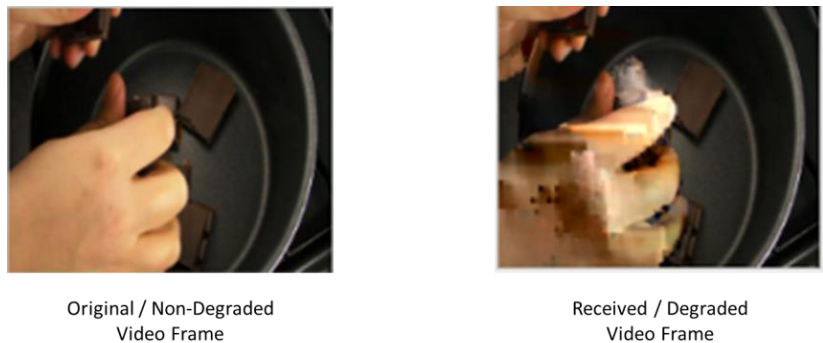

Figure 2. Sample Original and Degraded Video Frame

4. **Proposed Method**

Quality of service should be a primary focus for a video streaming service provider. The network strength for streaming video may vary for each end user and hence quality of the video may degrade in the receiver side based on network parameter. Hence to provide a good quality of service to the end customer, we are proposing vQoS-GAN (video Quality of Service GAN) which can determine the multi label network state parameters and reconstruct the degraded video to its original form. The solution to identify the degree of degradation from the received video data in the receiver side and to restore or reconstruct the degraded data as close to its original form is a highly difficult task. The proposed two different solution output can be solved using complex deep learning algorithm called Generative adversarial network (GAN). There is one important reason to implement the solution with GAN. There are many deep learning architectures that would be able to estimate the network state parameters. The estimation is possible only by comparing both the original and the received degraded video together. But in real application scenario, the original video data will not be available. The estimation of network parameters should be performed only with the received degraded video data. The solution proposes a new and unique design of generative adversarial network which can predict the network state like data rate and packet loss rate from the degraded video data and restore the degraded data to its original form based on the network state estimation.

Generative Adversarial network is a popular algorithm and marked a remarkable change in growth among the deep learning based industrial world. There are many variants of generative adversarial network based on the application need. But the initial aim of generative adversarial network was to create more synthetic images which never existed in the real world before. Generative adversarial network holds two different neural networks namely Generator network and the Discriminator network. Generator network tries to generate or create a new image whereas the discriminator network judges whether the generated image from the generator is a real one or a fake one. In general, the generative adversarial network is an unsupervised learning algorithm. So, for our solution, the basic design of generative adversarial network should be modified to a semi supervised learning method to feed the label parameters like data rate and packet loss rate along with the input images. Initially the generator does not have any knowledge regarding how a real or original non degraded image will look like. Similarly, the discriminator has good idea about real or original non degraded image but does not have knowledge regarding how a fake or received degraded image will look like. Both the network shares their knowledge to each other to learn their unknown parameters with respect to given input degradation level labels.

**4.1 Generator Network**

The Generator network is fed with the received degraded image along with the label parameters such as data rate and packet loss rate. The goal of the generator is to reconstruct or restore the input degraded image to its original image form. For achieving this goal, neural network layers in the generator are designed with the capability of performing image reconstruction. There are a few neural network architectures that can perform image reconstruction. The most popular and robust among them is the Autoencoder architecture. Two important modules in Autoencoder are 'Encoder' and 'Decoder'. The input image data passes to the encoder module first. The output of the encoder is the encoded data of the input image. The encoded output is also called as the latent representation of data. Network layers in the encoder learns the complex representation by compressing the important features of the actual input image. These encoded values are then passed to the decoder module. Decoder plays a critical role by regenerating or reconstructing the encoded data to its actual original input image form. From the compressed features of the input image, the decoder learns to restore the encoded data by decoding it appropriately. Although the image reconstruction process done by decoder is lossy, it performs the restoration of image data to a great extent. The architecture flow of generator network is shown in Figure 4. The Encoder performs down sampling operation with 3 convolutional layers and the final encoded output has 4000 dimensions of latent vectors. Decoder performs up sampling operation by taking transpose of 3 convolutional layers and finally outputs reconstructed image features.

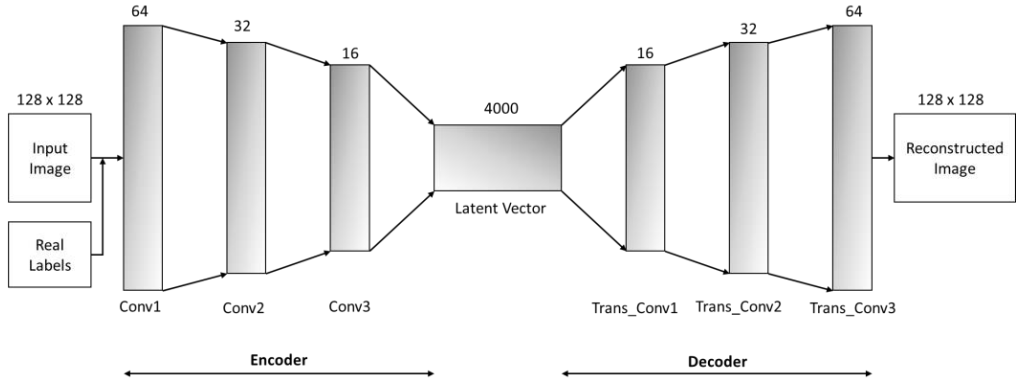

Figure 4. Generator Architecture of vQoS-GAN

With the autoencoder architecture in the generator network, the degraded image attained in the receiver side of the network can be restored to its original non degraded image form. The level or degree of image degradation varies for each received data and it is based on network parameters like data rate and packet loss rate. The perfect restoration of image is not possible without these parameters. With the knowledge of network state parameters, the model can restore the degraded image in a smoother way. There may be many neural network algorithms used for restoring the degraded image. The input dataset used to solve these existing techniques would have some constant level of degradation compared to its actual input image. Hence the model for image restoration could be trained with this predefined constant level of degradation. But in a real scenario, the degradation level attained in the receiver side varies dynamically. The existing algorithm will not be able to reconstruct the degraded image perfectly since the model does not have knowledge about how much percentage of reconstruction need to be performed for the given input image. Our proposed algorithm will guide the model with necessary labels such as data rate and packet loss rate which indicate the amount of reconstruction to be performed for the given input image. The data rate and packet loss rate are input labels for the model. Both degraded image and its corresponding label are given as input to the generator network. At each iteration, the generator network will try to create a restored or reconstructed output image for the given degraded input image.

## 4.2 Discriminator Network

Discriminator network is another important module in generative adversarial network architecture. Discriminator teaches generator network for differentiating whether the generated output is real or fake. Real image refers to expected desired output and fake image refers to incorrect output. Discriminator network is designed with series of convolutional layers capable of judging the generated output image. Discriminator has knowledge of original input image without any degradation. The network learns how the generated fake image from generator differs from the original image. Since the discriminator does not have any prior knowledge about fake images, the network start learning it and keeps adjusting its weight parameter accordingly by recording that the generator has produced a fake image. From the judgement of discriminator network, the generator start learning about the fake images and fine tunes its weight parameter towards generating real images in every iteration. Additionally, the discriminator network architecture is designed to predict the label parameter such as data rate and packet loss rate. The architecture flow of Discriminator network is shown in Figure 5. Discriminator network follows a usual CNN network with series of 4 convolutional layers by taking image as an input feature. The final layer follows with 3 different fully connected layers for predicting the data rate label, packet loss label and validating real or fake images.

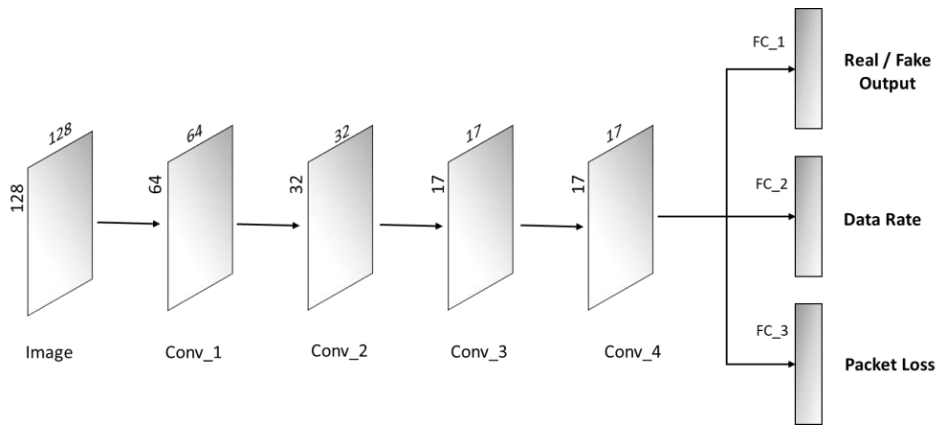

Figure 5. Discriminator Architecture of vQoS-GAN

There are few GAN architectures which supports for generating images with respect to specific given labels in a semi supervised manner. Most of the prediction based on input labels are done for multi class only. Multi class classification represent predicting single class from a single input image. i.e., each input image represents one class label only. Classifying more than one class in a single image data represent multi label class prediction. In the given problem statement, prediction should be done for data rate and packet loss label for each input video frames. The proposed architecture has been designed to support multi label classification by predicting both data rate and packet loss rate. The overall architecture flow for the proposed solution and the model training process can be seen in the Figure 6.

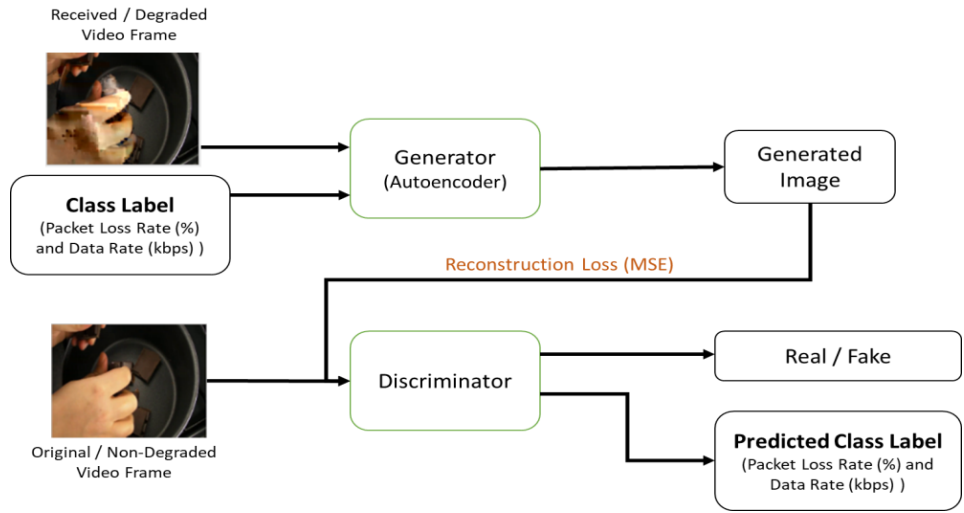

**Figure 6. Proposed vQoS-GAN Model Architecture**

The model is trained on learning the reconstruction procedure for each input degraded image based on the multi label parameters provided to restore the image closed to the actual non degraded original image. Once the model has been trained, the degraded image alone can be feed to the trained model as input. The Discriminator will predict the network state parameters such as data rate and packet loss rate for that input degraded image. The Generator network will provide the restored image output for the given degraded image. The process of predicting the model output is shown in Figure 7.

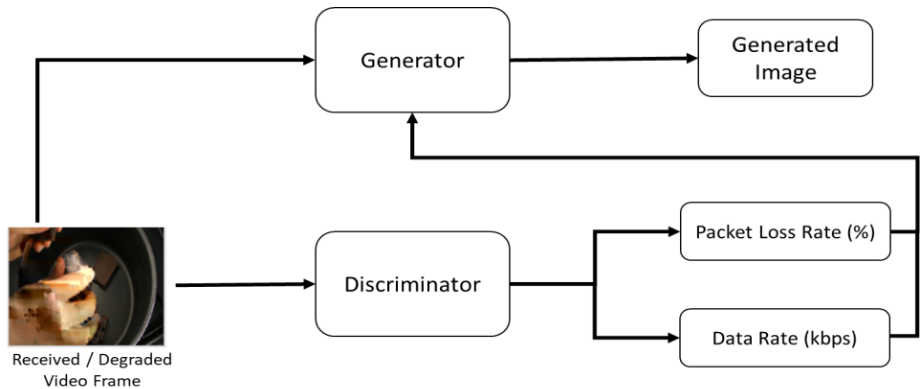

**Figure 7. vQoS-GAN Model Prediction Process**

With the proposed design, both network state estimation and restored non degraded image can be retrieved. This solution helps the telecom or content-based operator to provide a good service to the customer with improvised video quality and to enrich the end user experience.

The proposed architecture code flow has been described. Here we can see the learning of model performed for both generator and discriminator network along with multi label input parameters.

**Algorithm: vQoS-GAN model Execution Process**

**Generator**:

> Input Image: $I_g \in \{I_{recv}, Label_{real}\}$
>
> Output Image: $I_{recon} \in \{Gen.Model\}$

**Discriminator**:

> Input Image: $I_d \in \{I_{org}, I_{recon}\}$
>
> Output: $O_{res} \in \{Val(R,F), P_{Label}\}$

**Combined Process**:

for i = 0 to i = max_epoch:

  load receive image $I_{recv} \in \{R0 \text{ to } Rn\}$

  load original image $I_{org} \in \{I0 \text{ to } In\}$

  assign real label $Label_{real} \Leftarrow \{Idx: 0 \text{ to } n\}$

  assign fake label $Label_{fake} \Leftarrow \{Idx: 0 \text{ to } n\}$

  train generator $Gen_{model} = \{I_g, I_{recon}\}$

  train discriminator $Disc_{model} = \{I_d, O_{res}\}$

  calculate generator and discriminator losses

  $Disc_{loss} = Mean\{Disc_{real}, Disc_{fake}\}$

$$Gen_{loss} = Combined(I_{org},\ I_{recv}, label)$$

if (save the model intermediate weights):

   save_generator_model()

   save_discriminator_model()

## 5. Result and Discussion

The proposed deep learning semi supervised generative adversarial network architecture can estimate the network state data rate and packet loss from the given received degraded image. The Hardware configuration used for training the model is NVIDIA GTX 1080 8GB GPU with 16 GB RAM. Average of 10 epochs is trained for per hour duration. During model training, both original non degraded video data and its corresponding received degraded video data with class label data rate and packet loss are given to the network. Once the model is trained, the network state parameters data rate and packet loss can be predicted by feeding degraded video data to the discriminator network. Additionally, the degraded video data can be restored to its original form by feeding it to the generator network along with predicted network state labels. The model is trained for data rate 1200 kbps and 1600 kbps with packet loss of 0.05 %, 0.1 % and 0.25 %. The resultant prediction result for each class label in shown in Figure 8 with confusion matrix format.

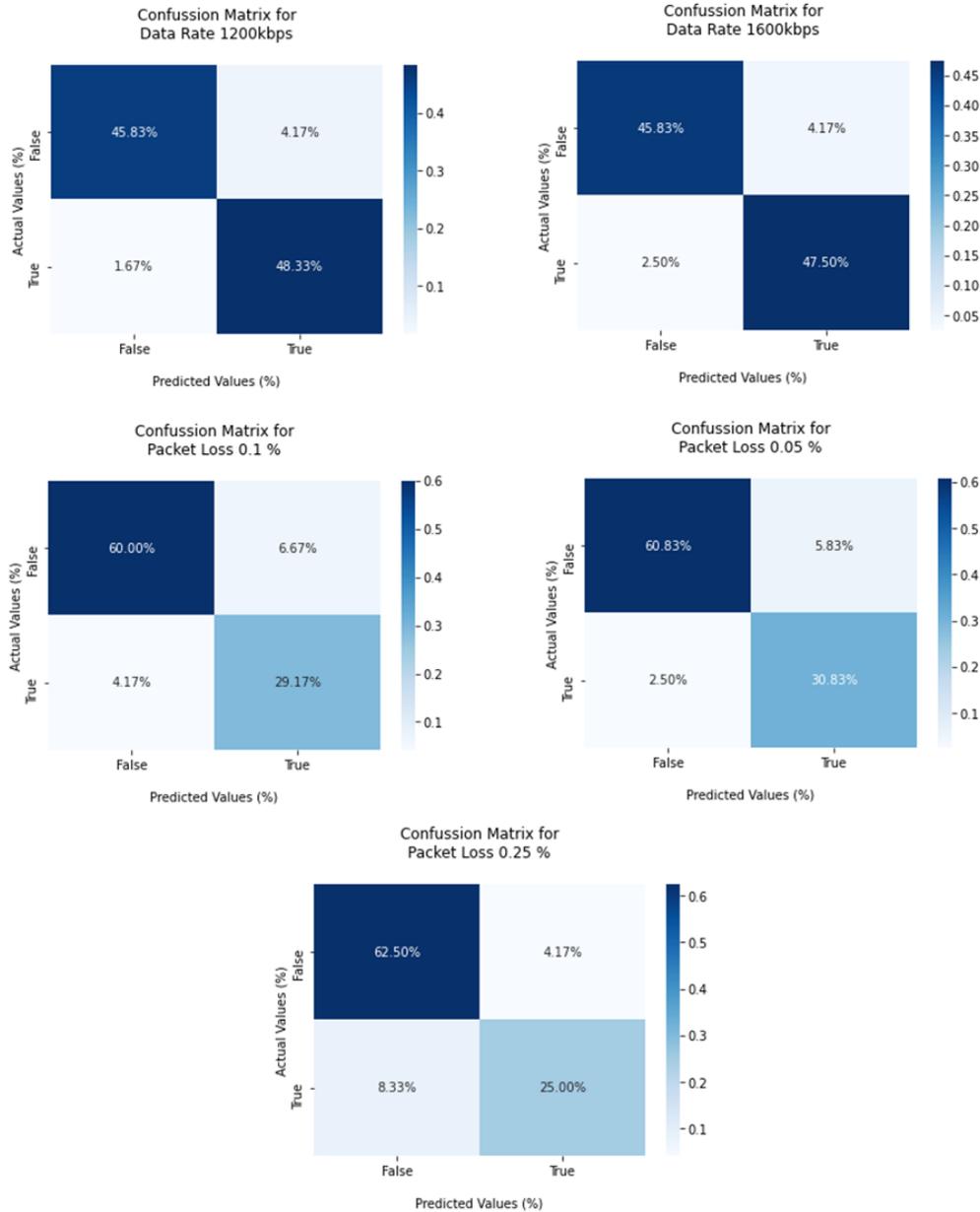

**Figure 8. Confusion Matrix for the model performance**

The proposed model architecture has been trained and compared with another model built with custom convolutional neural network. A section of video data has been clipped for both original and received video data from the actual large video dataset and the model has been trained for that clipped frames from the video. A custom CNN model is developed for estimating network state data rate and packet loss. But the CNN model needs both original and received image as input to predict the network state. The proposed GAN model needs only the received image or predicting the network state data rate and packet loss. From the Table 1. Accuracy comparison with proposed models, we can see the proposed GAN model predicts network state with higher accuracy than the custom CNN model.

| Model built for network state estimation | Accuracy score |
|---|---|
| Custom CNN model | 82 % |
| Proposed GAN model | 91% |

Table 1. Accuracy comparison with proposed models

The degraded video frame along with predicted network state labels from the discriminator output are given as input to the generator network. The generator network performs frame reconstruction and restore the input degraded video frame to its original form. The output of the generator module provides the reconstructed video frame with reduced level of degradation. The encoding decoding architecture in the generator network can restore the degraded image and provide output with less amount of distortion. The sample reconstructed output video frame can be seen in Figure 9.

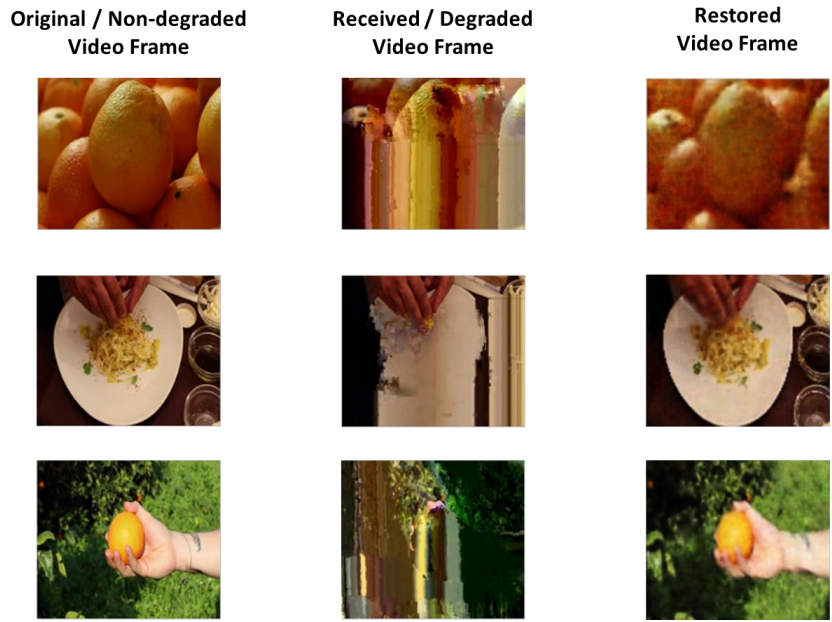

Figure 9. Reconstructed video frame from degraded image

A standalone python user interface application has been developed and tested with different data rate and packet loss. The results achieved with the trained model can be seen in Figure 10. The degraded video with network state 1600 kbps data rate and 0.25 % packet loss is detected correctly and the degraded received video is reconstructed to its original form for a good user experience. Another video of network state 1200 kbps data rate and 0.1 % packet loss is detected correctly, and the degraded received video is restored.

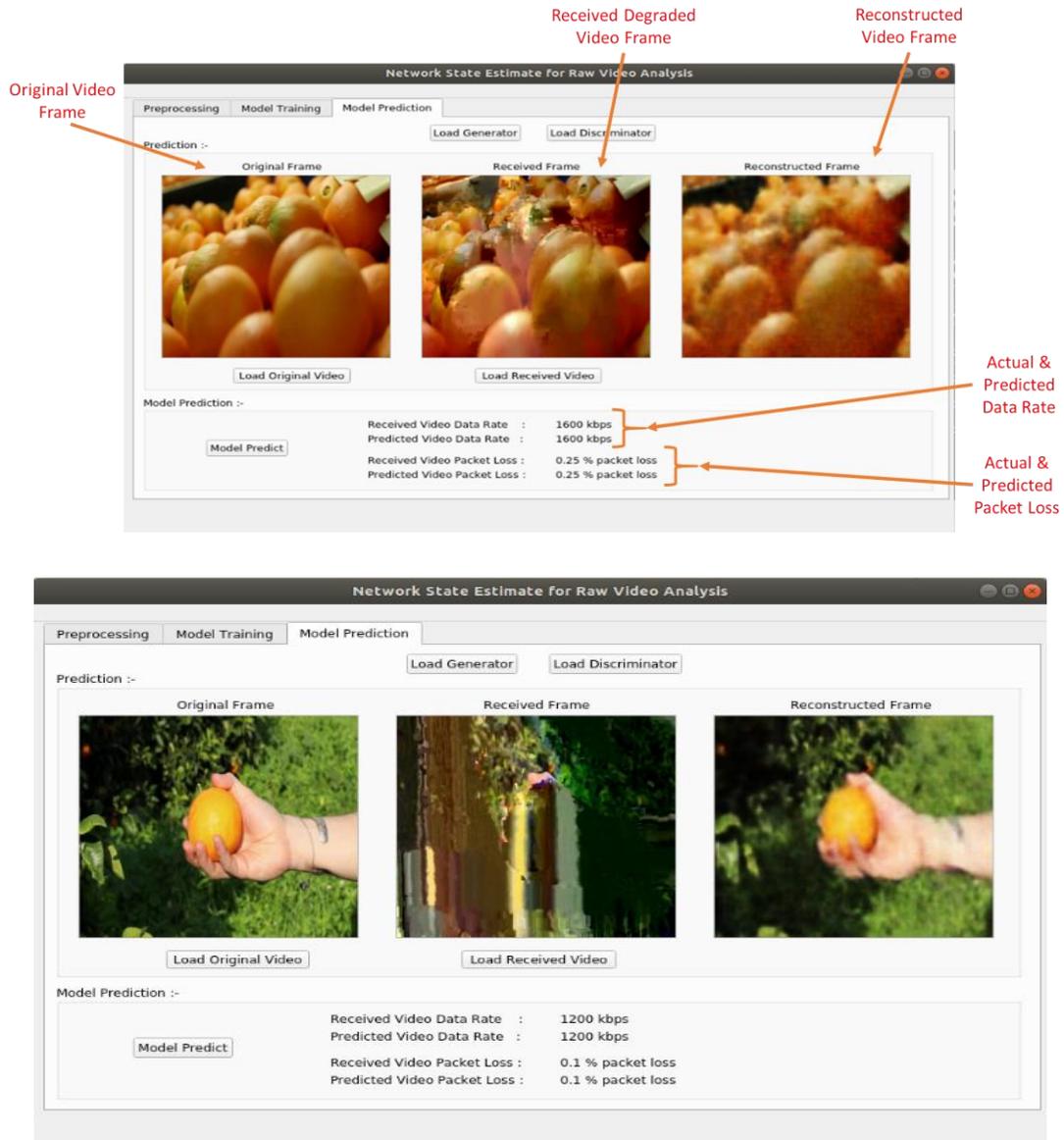

**Figure 10. The Application Module that was developed**

## 6. Conclusion and Future Work

Huge demand of video streaming services leads to more network traffic congestion. Network traffic is monitored using specialized hardware probes. In a practical scenario, it may not be possible to tap into the network at different points to monitor the network traffic congestion. Service providers need to serve their consumer with good quality video at the receiver or end user side. The quality of the video at the receiver side seems to be degraded with respect to network state parameters such as data rate and packet loss.

Network state estimation is normally performed by feeding both original and received video frame to the deep learning model. But in real application scenario, it is difficult to gather original video frame data. The proposed generative adversarial network model predicts the network state namely data rate and packet loss from the degraded video frame alone. The proposed algorithm gives good accuracy and seems more efficient compared to normal custom convolutional neural network.

The proposed solution has the following unique aspects:

• Deep learning-based GAN (Generative Adversarial Network) estimates network state and additionally generates good quality video from lossy video

• The design can be scaled with minimal modifications to estimate any other additional network state parameters like packet latency, jitter etc.

## 7. REFERENCES

[1] A. Nahli, S. Cao, Z. Jia, R. Ma and S. Xu, "Dataset and Network Structure: Towards Frames Selection for Fast Video Deblurring," in IEEE Access, vol. 9, pp. 61369-61382, 2021, doi: 10.1109/ACCESS.2021.3074199.

[2] S. Zhang, P. Li, Y. Meng, L. Li, Q. Zhou and X. Fu, "A Video Deblurring Algorithm Based on Motion Vector and An Encorder-Decoder Network," in IEEE Access, vol. 7, pp. 86778-86788, 2019, doi: 10.1109/ACCESS.2019.2923759.

[3] C. -H. Liang, Y. -A. Chen, Y. -C. Liu and W. Hsu, "Raw Image Deblurring," in IEEE Transactions on Multimedia, doi: 10.1109/TMM.2020.3045303

[4] H. Kim, D. Seo, J. Jung, D. Cha and D. Lee, "Blind Motion Deblurring for Satellite Image using Convolutional Neural Network," 2019 Digital Image Computing: Techniques and Applications (DICTA), 2019, pp. 1-8, doi: 10.1109/DICTA47822.2019.8945977.

[5] Z. Jiang, Y. Zhang, D. Zou, J. Ren, J. Lv and Y. Liu, "Learning Event-Based Motion Deblurring," 2020 IEEE/CVF Conference on Computer Vision and Pattern Recognition (CVPR), 2020, pp. 3317-3326, doi: 10.1109/CVPR42600.2020.00338.

[6] S. Su, M. Delbracio, J. Wang, G. Sapiro, W. Heidrich and O. Wang, "Deep Video Deblurring for Hand-Held Cameras," 2017 IEEE Conference on Computer Vision and Pattern Recognition (CVPR), 2017, pp. 237-246, doi: 10.1109/CVPR.2017.33.

[7] J. Tong, X. Wu, D. Ding, Z. Zhu and Z. Liu, "Learning-Based Multi-Frame Video Quality Enhancement," 2019 IEEE International Conference on Image Processing (ICIP), 2019, pp. 929-933, doi: 10.1109/ICIP.2019.8803786.

[8] H. Huang, I. Schiopu and A. Munteanu, "Frame-Wise CNN-Based Filtering for Intra-Frame Quality Enhancement of HEVC Videos," in IEEE Transactions on Circuits and Systems for Video Technology, vol. 31, no. 6, pp. 2100-2113, June 2021, doi: 10.1109/TCSVT.2020.3018230.

[9] A. Khasanova, A. Makhmutova and I. Anikin, "Image Denoising for Video Surveillance Cameras Based on Deep Learning Techniques," 2021 International Conference on Industrial Engineering, Applications and Manufacturing (ICIEAM), 2021, pp. 713-718, doi: 10.1109/ICIEAM51226.2021.9446438.

## 8. Author Biography

| Photo | Biography |
|---|---|
| 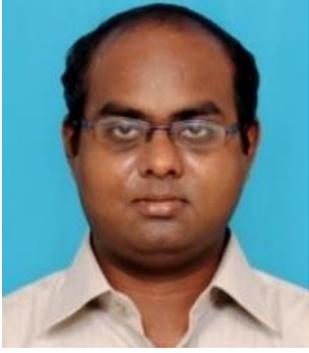 | **Renith G** received his B.E. Degree in Electronics and Communication Engineering from Anna University, India, in 2010 and received M.Tech Degree in Computer science and Engineering from SRM Institute of Science and Technology, India, in 2020. He works with HCL Technologies, India as Technical Lead. His current research interests include Image Processing, Computer Vision and Deep Learning. |
| 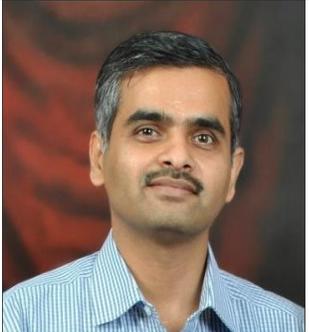 | **Harikrishna Warrier** received his B-Tech in Electronics and Electrical Communications from IIT Kharagpur and has done Post Graduate Certification in Business Management from XLRI Jamshedpur. He works with HCL Technologies, India as Solutions Director. His current research interests include MLOps, Data Centric AI, Edge Analytics and application of AI/ML in 5G. Hari has filed 7 patents in the areas of analytics and wireless communications and has 2 publications in international conferences. |
| 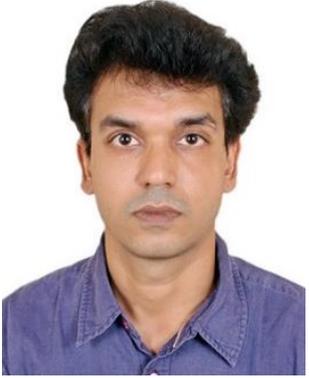 | **Yogesh Gupta** received his Master's degree in Computer Science from IP University, Delhi, India. He works with HCL Technologies, India as a Global Technology Director. His current research interests include MLOps, Federated Learning, Explainable AI and Lake-house architectures. Yogesh has filed 15+ patents in analytics, process automation and cloud areas. |